4DTIP of MEMS & MOEMS — Stresa, Italy, 26-28 April 2006# 3-D SELF-ASSEMBLED SOI MEMS: AN EXAMPLE OF MULTIPHYSICS SIMULATION

*C. Méndez\*, C. Louis\*, S. Paquay\*, P. De Vincenzo\*, I. Klapka\*, V. Rochus†, F. Iker‡, N. André‡ and J.-P. Raskin‡*

\*Open Engineering, rue des Chasseurs Ardennais 8, B-4031 Angleur (Liège), Belgium
†Université de Liège, Bâtiment B52, Chemin des Chevreuils 1, B-4000 Liège, Belgium
‡Université catholique de Louvain, Place du Levant 3, B-1348 Louvain-la-Neuve, Belgium**ABSTRACT**

MEMS devices are typical systems where multiphysics simulations are unavoidable. In this work, we present possible applications of 3-D self-assembled SOI (Silicon-on-Insulator) MEMS such as, for instance, thermal actuators and flow sensors. The numerical simulations of these microsystems are presented. Structural and thermal parts have to be strongly coupled for correctly describing the fabrication process and for simulating the behavior of these 3-D SOI MEMS.## 1. INTRODUCTION

The main challenge of a numerical simulation of MEMS devices is the coupling between several scientific fields that take place in these structures (mechanical, electrical, thermal, fluidic, etc.). Frequently, as in the cases that we present here, a strong coupling between the fields is essential for a good description of these systems.

In this work, we present the modeling of the fabrication process and the operation of 3-D self-assembled SOI (Silicon-on-Insulator) MEMS using the multiphysics code named Oofelie [1]. The out-of-plane 3-D characteristic of these MEMS gives the possibility for several applications such as thermal actuators and flow sensors shown in Fig. 1. The SOI technology employed in the fabrication allows an almost direct integration with SOI MOS circuits, and it is worth to note that these circuits have a better behavior than traditional bulk silicon technology concerning power consumption, resistance to irradiation and high temperature operation [2].

These MEMS are multilayered cantilevers composed of Si, $Si_3N_4$ and Al. As the materials are deposited at high temperatures and because they have different thermal expansion coefficients, when temperature decreases down to room temperature the structure is under thermal stress and this stress leads to the self-assembling of the device when it is released.

The final shape of the structure can be controlled by modifying the number of layers, the thickness of each of them or making an additional thermal process after the release. In this way, rather flat or very curved structures can be obtained. These different possibilities were simulated numerically and comparisons with laser interferometry measurements were made. In order to obtain good results, the simulation has to take into account the strong coupling of mechanical and thermal fields and the large displacements that can take place in some cases.

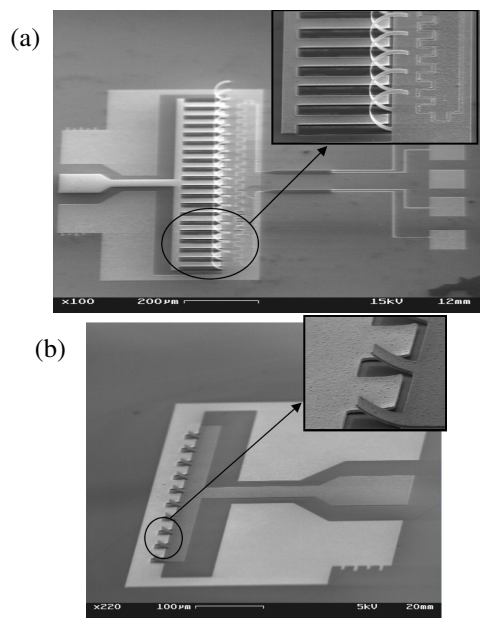

Figure 1: 3-D self-assembled MEMS: (a) thermal actuator, (b) flow sensor.

After the 3-D shape is obtained, it can be used as an actuator or a sensor. For example, adding a microheater at the anchor of the cantilever the temperature of the whole structure can be changed. The microheater can be made of doped polysilicon, and a DC current flowing through it increases the temperature by Joule effect. As a consequence, the thermal stress that bends up the structure is changed and therefore its shape is modified as well. In this case, besides the thermal and mechanical fields, we have to add the electrical part to the simulations.

©TIMA Editions/DTIP 2006     ISBN: 2-916187-03-0

When a gas or liquid flow is flowing parallel to the wafer surface, a pressure is applied along the cantilevers composing the 3-D MEMS flow sensors. This gas or liquid pressure will push down to the surface the cantilevers. Including a piezoelectric material at the cantilever anchor, stress variations due to the change of its shape can be detected and thus it is possible to quantify the intensity of the flow. Therefore, in this other application, a new class of materials has to be included in the simulation.

A description of the fabrication process is given in Section 2. In Section 3, we show how the fabrication and operation of these MEMS can be simulated numerically and conclusions are given in Section 4.

## 2. FABRICATION

The fabrication of these 3-D MEMS is fully compatible with a standard process CMOS. Over a SOI wafer, a LPCVD (Low Pressure Chemical Vapor Deposition) $Si_3N_4$ layer is deposited at 800°C. At 150°C, a layer of Al is evaporated above the $Si_3N_4$ layer. A unique photolithographic step is made to define the pattern of the cantilevers and then afterwards etch the layers of Si, $Si_3N_4$ and Al. The last fabrication step is the release of the microstructures by the wet etching of the SOI wafer buried oxide in a mixture of HF and isopropanol [3].

As the thermal expansion coefficients of the layers which composed the 3-D MEMS are different, when the temperature decreases down to room temperature a thermal stress is generated. This stress is responsible for the self-assembling of the structure, and the resulting shape can be very different depending on the number of layers, the thickness of each of them, and if a thermal treatment is made after the release (Fig. 2).

In order to have a better understanding of the phenomena involved, several test structures were built. In Fig. 2a, released Si beams are shown and, as it could have been foreseen, they remain plane after the release. This experiment clearly shows that the thin silicon active layer from SOI wafers is free of stress gradient throughout its thickness. However, when the structures consist of two layers - Si and $Si_3N_4$ - the beams bend up (Fig. 2b). This is because the $Si_3N_4$ has a thermal expansion coefficient larger than the Si and thus the upper part of the structure tends to contract more than the lower part. In the trilayer case, the Al has the largest thermal expansion coefficient; however, the resulting shape is rather flat (Fig. 2c). The reason for this is that, in addition to the thermal coefficient, deposition temperature and thickness of each layer have to be considered. The Al layer is deposited at a lower temperature (150°C) than the $Si_3N_4$ layer (800°C), and thus, the difference between the thermal expansion coefficients is compensated.

The thickness of each layer is also important, because in this case, the Al layer is very thick (900 nm) compare with the other two layers (100 and 250 nm for Si and $Si_3N_4$, respectively). As a result, the Si and $Si_3N_4$ located at the bottom part of the cantilevers contract, balance the Al contraction, and then the structure remains approximately flat after the release at room temperature. Nevertheless, as can be seen in Fig. 2d, after an additional thermal process, the trilayer can adopt a very curved form. The thermal treatment can be a rapid annealing (2 min at 600°C), or a longer annealing at a lower temperature (30 min at 432°C).

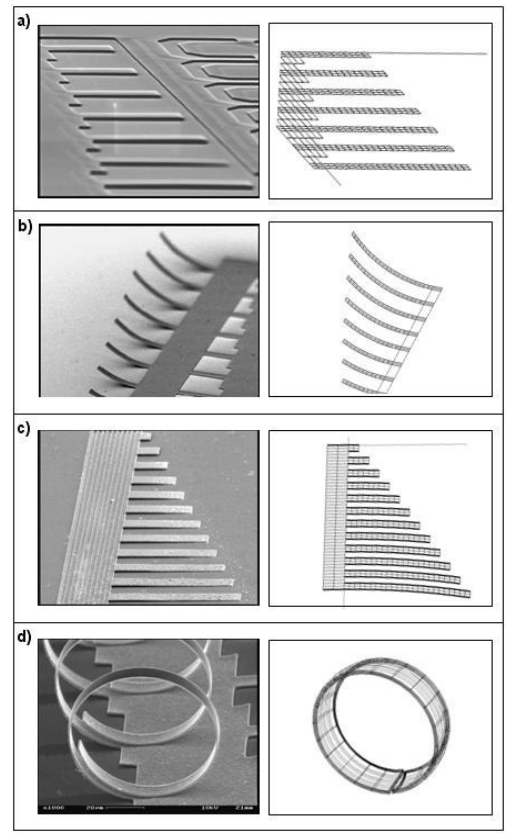

Figure 2: Different shapes after the fabrication of multilayered structures: (a) Si monolayer, (b) Si-$Si_3N_4$ bilayer, (c) Si-$Si_3N_4$-Al trilayer, (d) Si-$Si_3N_4$-Al after a thermal treatment.

## 3. NUMERICAL SIMULATION

### 3.1. Fabrication

Concerning the simulation of the fabrication process of this kind of MEMS using a finite element code, two major characteristics have to be taken into account. One of them is the strong coupling between the structural and the thermal parts, and the other is the large displacement that these structures can present.

Because of these features, good results can only be obtained if a geometrically nonlinear analysis with a strong coupling between thermal and mechanical fields is considered. Specifically, to obtain convergence, the thermal contribution has to be taken into account in the calculation of the tangent stiffness matrix, and therefore has to be considered as an internal force, not merely as perturbation. Although the cases presented here are simple, in order to allow the simulation of more complex geometries without changes, second order tridimensional elements were used.

A comparison with experimental results was made for the bilayer case. For this, the shape of several cantilevers of different lengths was measured using laser interferometry. The profiles are shown in Fig. 3, and the fit using a circular function gives practically the same radius of curvature for all the samples.

The simulation was made using the parameters of Table 1 and the radius obtained was R = 47.37 μm, in agreement with analytical predictions [4]. This indicates that, although other sources of stress may be present (e.g. lattice mismatch at the interface), the bending of the structure can be accurately simulated considering only thermal effects.

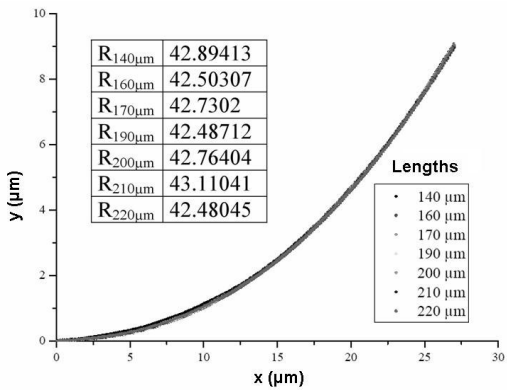

Figure 3: Shapes of bilayered cantilevers of various lengths and fitted radii of curvature for each case.

Table 1: Main characteristics of the bilayers (Si-$Si_3N_4$) presented in Fig. 3.

|  | 1st layer: Si | 2nd layer: $Si_3N_4$ |
|---|---|---|
| Young's modulus (GPa) | 130.0 | 270.0 |
| Poisson's ratio | 0.279 | 0.270 |
| Thermal exp. coeff.(1/°C) | 2.33E-6 | 6.06E-6 |
| Thickness (nm) | 100 | 100 |
| Deposition temp. (°C) | - | 800 |

In the above case, the curvature is not the maximum curvature that can be obtained for this given bilayer Si-$Si_3N_4$. For a fixed Si layer thickness of 100 nm, the deflection of the cantilever tip is plotted in Fig. 4 versus the $Si_3N_4$ thickness. We can see that the maximum deflection for a fixed silicon film of 100 nm-thick is reached with a $Si_3N_4$ layer of approximately 35 nm. The limits where the $Si_3N_4$ thickness tends to zero and infinity, correspond to monolayers of Si and $Si_3N_4$, respectively, and the curvature goes to zero.

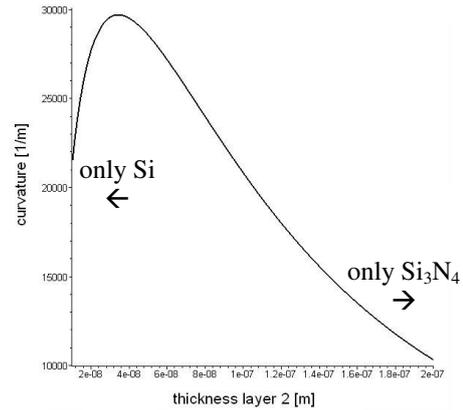

Figure 4: Bilayered cantilever curvature as a function of the $Si_3N_4$ layer thickness.

A similar analysis can be done in the trilayer case. If we fix again the thickness of the Si layer, we can plot the curvature as a function of the thicknesses of $Si_3N_4$ and Al as in Fig. 5. For each variable, the behavior is similar to that of Fig. 4, and as the Al has the higher thermal expansion coefficient, the absolute maximum occurs when the $Si_3N_4$ thickness is equal to zero.

Fixing Si and $Si_3N_4$ thicknesses at 100 and 50 nm, respectively, the maximum curvature corresponds to a thickness of Al of approximately 50 nm. It is not always possible to maximize the curvature of the trilayer, because some of the layers should have minimal thicknesses to preserve, for example, electrical properties. Nevertheless, this kind of analysis shows us that, to increase the cantilevers deflection, the materials that will contract more should be put in the geometrical upper part in order to maximize the bending momentum.

We must mention that the above analysis was made without considering the plasticity of the Al layer. Therefore, it is valid only if the materials of the trilayer are always in the elastic regime. In the case of Fig. 2d, for example, the main difficulty in the simulation is the treatment of the geometric nonlinearity, but to make quantitative comparisons with experiments the inclusion of the plasticity of the Al is unavoidable.

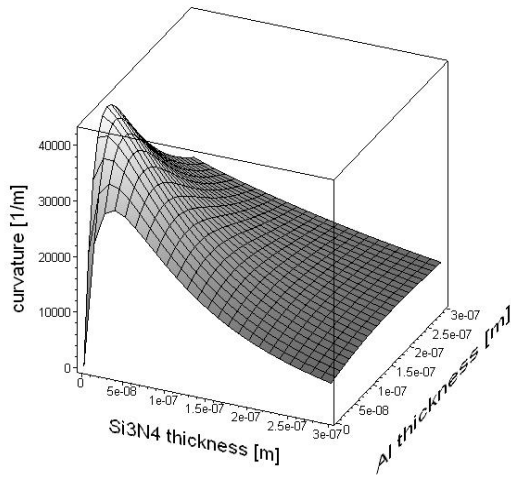

Figure 5: Trilayered cantilever curvature as a function of $Si_3N_4$ and Al thicknesses.

### 3.2. Operation

The 3-D out-of-plane characteristic of these kinds of MEMS allows its use in a variety of applications. We will show two examples; one of them is a thermal actuator and the other a flow sensor (Fig. 1).

In the thermal actuator, a microheater made of thin doped polysilicon is located at the anchor of the cantilever. Thus, by Joule effect, a tunable DC current can increase the temperature of the multilayer, and hence can change its shape. At room temperature (zero current) the structure has its maximum deflection and when the intensity of current is increased, the temperature rises, and the cantilever goes down towards the wafer surface. Hence, for the simulation of this thermal actuator, not only mechanical and thermal but also electrical effects have to be included. In Fig. 6, a detail of the microheater is shown and the temperature distribution can be seen.

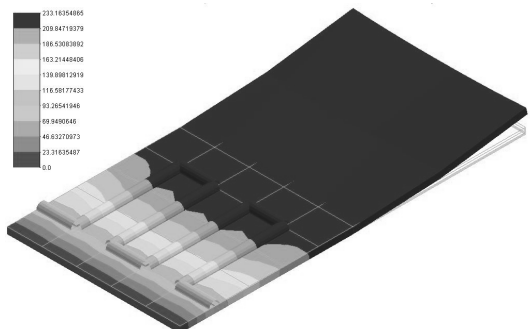

Figure 6: Detail of the microheater and temperature distribution at the base of the cantilever.

As a direct application, these thermal actuators can be used as switches, but they can also be used as components of more sophisticated devices. For example, in Fig. 7 it is shown how an array of thermal actuators can move an object placed on the top of the array. Turning on and off the current in the microheaters in the adequate sequence, the object can be displaced.

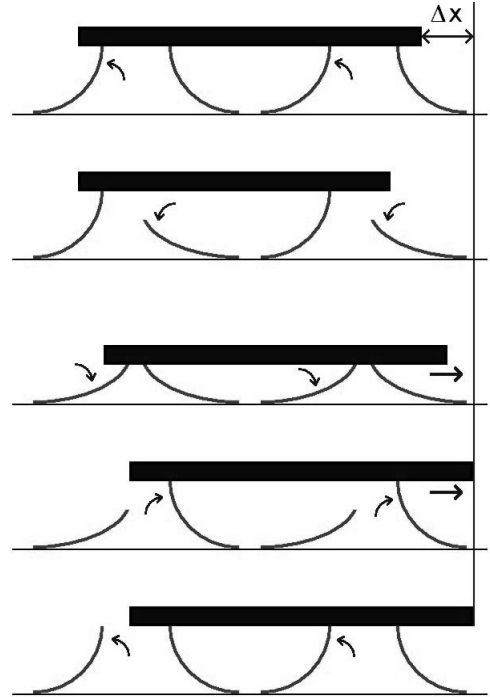

Figure 7: Example of application for the 3-D MEMS thermal actuators [5].

On a two dimensional array, even a rotational movement can be achieved if the appropriate control on the DC current is implemented.

In order to use these structures as flow sensors, we have to measure the deformation of the cantilevers when a gas or fluidic flow is applied to it. One alternative is to measure changes of capacitance between the multilayer and a fixed electrode lying on the substrate. Another one is to see how the stress inside the cantilever changes when a flow modifies its shape. This second possibility is to implement a piezoelectric material at the anchor of the cantilever, where the mechanical stress is maximum. Thus, the intensity and direction of the flow is measured through the variations of the potential generated in the piezoelectric. The simulation in this transducer has to include a new class of materials (piezoelectric) and preliminary results are shown in Fig. 8. When a pressure is applied on the multilayer, changes in the electric potential can be clearly observed.

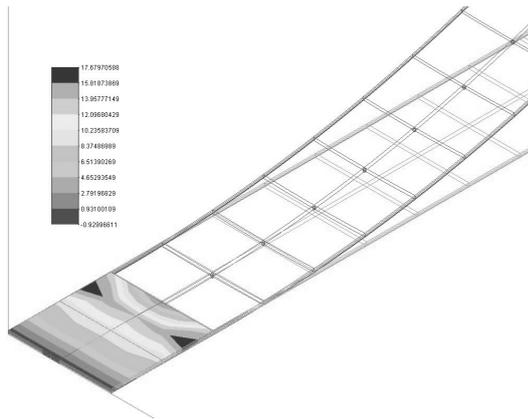

Figure 8: Detail of the piezoelectric material located at the anchor of the cantilever to sense the flow.

## 4. CONCLUSIONS

We have shown how out-of-plane 3-D MEMS can be obtained using the same fabrication procedures that for SOI MOS circuits. This full CMOS compatibility opens great opportunities for the fabrication of sensors and actuators with their associated electronics on the same silicon chip.

We have also seen that taken into account the properties of the materials used, different shapes can be obtained modifying the thicknesses of the layers or using additional thermal treatments. The 3-D nature of these systems permits their use in a variety of devices and we have presented a thermal actuator and a flow sensor as two potential applications.

For the numerical simulation of these MEMS we have used the finite element code Oofelie. Implementing a strong coupling between mechanical and thermal fields and making a geometrically nonlinear analysis, the major features of the various fabrication steps have been accurately simulated.

Due to the multiphysics characteristic of the code, the transducer (mechanical deformation to electrical signal) of these MEMS can also be incorporated. Thus, the electrical part of the thermal actuator and the piezoelectric material for the flow sensor were included in preliminary simulations.

A more quantitative comparison between simulations and experiments for these 3-D MEMS flow sensors and thermal actuators are under investigation.

## 5. ACKNOWLEDGMENTS

This work was supported by the Program Alban, the European Union Programme of High Level Scholarships for Latin America, scholarship no. E04E043398AR.